\documentstyle[twocolumn,aps,epsf]{revtex}

\newcommand{\beq}{\begin{equation}}
\newcommand{\eeq}{\end{equation}}
\newcommand{\ba}{\begin{array}{ccc}}
\newcommand{\ea}{\end{array}}

\begin{document}
\draft
\title{QCD-Like Behaviour of High-Temperature Confining Strings}
\author{M. C. Diamantini$^a$\cite{bylinea} and C. A.
Trugenberger$^b$}
\address{$^a$$^b$Theory Division, CERN,CH-1211 Geneva 23, Switzerland}
\address{$^b$InfoCodex SA, av. Louis-Casa\"i 18, CH-1209 Geneva, Switzerland}
\address{$^a$cristina.diamantinitrugenberger@cern.ch}
\address{$^b$ca.trugenberger@InfoCodex.com}
\date{\today}
\maketitle
\begin{abstract}
We show that, contrary to previous string models, the high-temperature behaviour of the recently proposed
confining strings reproduces exactly the correct large-$N$ QCD result, a {\it necessary}
condition for any string model of confinement.
\end{abstract}

\narrowtext
Although fundamental strings \cite{witten} can be quantized only in critical
dimensions, strings in four space-time dimensions are of great
interest since there is a large body of evidence \cite{polc1}, recently confirmed by
numerical tests \cite{caselle}, that they can describe the confining phase of
non-Abelian gauge theories. However, a consistent quantum theory describing these
strings has not yet been found: the simplest model, the Nambu--Goto string, can
be quantized only in space-time dimension $D$ = 26 or $D \leq 1$ because of the
conformal anomaly; it is inappropriate to describe the expected smooth strings
dual to QCD \cite{poly1}, since large Euclidean world-sheets are crumpled.
In the rigid string \cite{kleinert}, the marginal term proportional to
the square of the extrinsic curvature, introduced to cure this problem, turns
out to be infrared irrelevant and, thus, unable to prevent crumpling.

Both these models also fail to describe the correct high-temperature behaviour of
large-$N$ QCD \cite{polc2}. As shown in \cite{polc3},
the deconfining transition in QCD is due to the condensation of Wilson lines, and the
partition function of QCD flux tubes can be continued above the deconfining
transition; this high-temperature continuation can be evaluated
perturbatively. So, any string theory that is equivalent to QCD {\it must} reproduce
this behaviour. However, the Nambu--Goto action has the wrong temperature dependence,
while the rigid string has the correct high-temperature behaviour but with a
wrong sign and an imaginary part signalling a world-sheet
instability \cite{polc2}.
At low temperatures, the behaviour of the rigid string was studied in \cite{klei1}.

Recently, two new models have been proposed: a first one, 
the confining string \cite{pima}, is based on an induced string
action explicitly derivable for compact QED \cite{pimo} and for
Abelian-projected SU(2) \cite{anto}; a second one, originally proposed in
\cite{poly2}, is based on a five-dimensional, curved space-time string action with
the quarks living on a four-dimensional horizon \cite{alvarez}. The interrelation between these
two models has been analysed in \cite{ellis}.

The confining string action possesses, in its world-sheet formulation, a non-local
action with a negative stiffness \cite{pimo,klei} that can be expressed as a
derivative expansion of the interaction between surface elements.
To perform an analytic analysis of the geometric properties of these
strings, this expansion must be truncated: this clearly makes the model
non-unitary, but in a spurious way. Moreover, since the stiffness is negative, a
stable truncation must, at least, include a sixth-order term in the derivatives
\cite{cris1}. In \cite{cris1,cris2} it was shown that, in the large-$D$
approximation, this model has an infrared fixed point at zero stiffness, corresponding to a
tensionless smooth string whose world-sheet has Hausdorff dimension 2, exactly
the desired properties to describe QCD flux tubes. As first
noticed in \cite{cris3}, the long-range orientational order in 
this model is due to an
antiferromagnetic interaction between normals to the surface, a mechanism
confirmed by numerical simulations \cite{chern}.
Moreover, it was shown in \cite{cris2} that this infared fixed point does not depend on the
truncation and is present for all ghost- and tachyon- free truncations, and that
the effective theory describing the infrared behaviour is a conformal field
theory with central charge $c = 1$.

In this paper we will study the high-temperature behaviour of the string  model defined
in \cite{cris1}. We will show that this model
has a high-temperature behaviour that agrees  in temperature
dependence,  {\it  sign and reality properties}
with the large-$N$ QCD result \cite{polc2}.
This result depends entirely on the higher order term and is totally independent of the stiffness. 
Finite-temperature confining strings
in (2+1) dimensions and in the presence of D0-branes have been studied in \cite{gerald}.

In Euclidean space, the action proposed in \cite{cris1} is:
\beq
S = \int d^2 \xi \sqrt{g}  g^{ab} {\cal D}_a x_\mu \left( t - s{\cal D}^2 +
 {1 \over M^2} {\cal D}^4 \right){\cal D}_b x_\mu  \ , \label{model}
 \eeq
where ${\cal D}_a$ are covariant derivatives with respect to the induced metric
$g_{ab} = \partial_a x_\mu \partial_b x_\mu$ on the surface ${\bf x}(\xi_0,
\xi_1)$.
The first term in the bracket provides a bare surface tension $2t$, while the
second accounts for the rigidity, with a stiffness parameter $s$ that we set
to its fixed-point value $s = 0$. In the third term, $M$ is a new mass scale. Since this term contains
the square of the gradient of the extrinsic curvature matrices, it suppresses the
formation of spikes on the world-sheet. In the large-$D$ approximation it
generates a string tension proportional to $M^2$, which takes control of the
fluctuations where the orientational correlation dies off.
To perform the large-$D$ analysis we introduce  a Lagrange multiplier \cite{david}
$ \lambda^{ab}$  that forces the induced metric  
$\partial_a x_\mu \partial_b x_\mu$ to be equal to the intrinsic metric $g_{ab}$, extending the
action (\ref{model}) to:
\beq
S\rightarrow S + \int d^2 \xi \sqrt{g} \left[\lambda^{ab} (\partial_a x_\mu \partial_b x_\mu - g_{ab}
)  \right] \ . \label{nmodel} 
\eeq

We parametrize the world-sheet in a Gauss map by $x_\mu (\xi) = (\xi_0, \xi_1,
\phi^i(\xi)),\ i = 2,...,D-2$. The value of the periodic coordinate $\xi_0$ is
$-\beta/2 \leq \xi_0 \leq \beta/2$ with $\beta = 1/T$ and $T$  the
temperature.
Note that  at high temperatures ($\beta \ll 1$), the scale $M^2$ can be temperature-dependent. 
This is not unusual in closed string theory as  has been shown by Atick and Witten \cite{witten1}.
The value of $\xi_1$  is
$-R/2 \leq \xi_1 \leq R/2$; $\phi^i(\xi)$ describe the $D-2$ transverse
fluctuations. We look for a saddle-point solution with a diagonal metric $g_{ab}
= {\rm diag}\ (\rho_0, \rho_1)$, and a Lagrange multiplier of the form
$\lambda^{ab} = {\rm diag}\ (\lambda_0/\rho_0, \lambda_1/\rho_1)$.
The action then becomes:
\begin{eqnarray}
S &&= S_0 + S_1\nonumber \\
S_0 &&=  A_{\rm ext} \sqrt{\rho_0 \rho_1} \left[ \right. t \left( {\rho_0 + \rho_1
\over \rho_0 \rho_1} \right) + \lambda_0 \left( { 1 -\rho_0 \over \rho_0}
 \right) \nonumber \\
&&+ \lambda_1 \left( { 1 -\rho_1 \over \rho_1} \right) \left. \right] \nonumber \ ,\\
S_1 &&= \int d^2 \xi \sqrt{g}  g^{ab}\partial_a \phi^i \left( t + {1 \over M^2} 
{\cal D}^4 \right) \partial_b \phi^i \ ,
\end{eqnarray}
where $\beta R = A_{\rm ext}$ is the extrinsic, projected area in coordinate
space, and $S_0$ is the tree-level contribution.
Integrating over the transverse fluctuations in the one-loop term $S_1$, we obtain, in the limit $R \to
\infty$:
\begin{eqnarray}
S_1 &&= {D - 2\over 2} R \sqrt{\rho_1} \sum_{n = - \infty}^{+ \infty} \int {d
p_1\over 2 \pi} \ln\left[ \right.  t \left( \omega_n^2 + p_1^2 \right) \nonumber \\
&&+\ p_1^2 \lambda_1 + \omega_n^2 \lambda_0 + {1\over M^2} 
\left( \omega_n^2 + p_1^2 \right)^3 \left. \right] \ ,
\end{eqnarray}
where $\omega_n = {2 \pi \over \beta \sqrt{\rho_0} } n$.
At high temperatures, satisfying
\beq 
(M^2 \beta^2) (t \beta^2) \ll 1 \ ,
\label{valid}
\eeq
the sixth-order term in the derivatives dominates in the one-loop term $S_1$
 when $ n \neq 0$. 
Using analytic regularization
$$
\int_{\rm reg} d x\ {\rm ln} (x^2 + a^2) = 2 \pi a \ ,
$$
and analytic continuation of the formula 
$$\sum_{n =
1}^\infty n^{-z} = \zeta(z) \ ,$$
 for the Riemann zeta function, with $\zeta(-1) = -
1/12$, we obtain for the $ n \neq 0$ contribution:
\begin{eqnarray}
&&{D - 2\over 2} R \sqrt{\rho_1} \sum_{n = - \infty}^{+ \infty} \int {d
p_1\over 2 \pi}\ {\rm ln} {\left( \omega_n^2 + p_1^2 \right)^3 \over M^2}\nonumber \\
&&= {D - 2\over 2}\sqrt{\rho_1 \over \rho_0} 12 \pi {R\over \beta} \sum_{n = 1}^{+ \infty}\sqrt{n^2}
= {D - 2\over 2}\sqrt{\rho_1 \over \rho_0} 12 \pi {R\over \beta} \zeta(-1)\nonumber \\
&&= -{D - 2\over 2}\sqrt{\rho_1 \over \rho_0}{ \pi R\over \beta}\ .
\end{eqnarray}
For $ n = 0$, instead, rewriting 
\begin{eqnarray}
&&{\rm ln}\left[ p_1^2(t + \lambda_1) + {1\over M^2} p_1^6\right] = {\rm ln} \left[{p_1^2 \over
M^2}  \left(M^2(t + \lambda_1) + p_1^4\right) \right] \nonumber \\
&&= {\rm ln}\left[ (p_1^2 + i
M\sqrt{\lambda_1 +t})\ (p_1^2 - i M\sqrt{\lambda_1 +t})\right] + {\rm ln} {p_1^2 \over
M^2}\ ,\nonumber
\end{eqnarray}
we obtain:
\begin{eqnarray}
&&{D - 2\over 2} R \sqrt{\rho_1}\int {d p_1\over 2 \pi}\ {\rm ln}\left[p_1^2(t + \lambda_1) 
+ {1\over M^2} p_1^6\right]\nonumber \\
&&= {D - 2\over 2} R \sqrt{\rho_1}\int {d p_1\over 2 \pi}\ 2 {\rm Re}\ {\rm ln}
(p_1^2 + i M\sqrt{\lambda_1 +t}) \nonumber \\
&&= {D - 2\over 2} R \sqrt{\rho_1}\ \sqrt{2 M} (\lambda_1 +t)^{1/4}\ .
\end{eqnarray}
The action $S = (S_0 + S_1)$ then becomes
\beq
S = S_0 + {D - 2 \over 2} R \sqrt{\rho_1} \left[ \sqrt{2 M} \left( \lambda_1 
+ t \right)^{1/4} - {\pi \over \beta \sqrt{\rho_0}} \right] \ .
\label{spa}
\eeq

The factor ${D-2\over 2}$ in $(S_0 + S_1)$ ensures that, for large $D$, the fields
 $\rho_0,\ \rho_1,\ \lambda_0$ and  $\lambda_1$ are extremal and  thus satisfy
the four-gap equations:
\begin{eqnarray}
&&{ 1 -\rho_0 \over \rho_0} =  0\ , \label{gap1}\\
&&{1 \over \rho_1} = 1 - {D - 2\over 2} { 1 \over 4 \beta } 
{\sqrt{2 M} \over (\lambda_1 + t )^{3/4} }\ , \label{gap2} \\
&&\left[ {1\over 2}(t - \lambda_1) + {1 \over 2\rho_1}(\lambda_1 + t ) - t -
\lambda_0 \right] + {D - 2\over 2} { \pi \over 2 \beta^2 } = 0 \ ,\label{gap3} \\
&&(t - \lambda_1) - {1 \over \rho_1}(\lambda_1 + t ) + 
\nonumber \\
&&{D - 2\over 2} { 1 \over  \beta } \left[ \sqrt{2 M} \left( \lambda_1 
+ t \right)^{1/4} - {\pi \over \beta^2 } \right] = 0 \label{gap4} \ .
\end{eqnarray}
Substituting (\ref{gap4}) into (\ref{spa}) we obtain a simplified form of the
effective action:
\beq
S^{\rm eff} = A_{\rm ext}  {\cal T} \sqrt{\rho_0 \over \rho_1}
\label{effac} \ ,
\eeq
with ${\cal T} =  2 (\lambda_1 + t )$  representing  the physical string
tension.

By inserting (\ref{gap2}) into (\ref{gap4}), we obtain an equation for 
$(\lambda_1 + t )$ alone:
\begin{eqnarray}
&&(\lambda_1 + t ) -  {D - 2\over 2} {5\over 8 \beta}\sqrt{2 M}
\left( \lambda_1 + t \right)^{1/4} \nonumber \\
&&+\ {D - 2\over 2} { \pi \over  2 \beta^2 } - t = 0 \ .
\label{quart}
\end{eqnarray}

Without loss of generality we  set 
\beq 
\left( \lambda_1 + t \right)^{1/4} = {\sqrt{2 M} \over \gamma} \ ,
\label{condiz}
\eeq
where $\gamma $ is a dimensionless parameter.
It is possible to show that, at high temperatures, when
\beq
t \beta^2 \ll {D - 2\over 2} \ ,
\label{cont}
\eeq
we can completely neglect $t$ in (\ref{quart}). Indeed, as we now show, $\lambda_1$ is proportional
to $(D-2)/\beta^2$. Note that this is compatible with the condition  (\ref{valid}) used before.
We can thus rewrite (\ref{quart}) as:
\beq
\lambda_1 -  {D - 2\over 2} {5\over 8 \beta}\gamma
\lambda_1^{1/2} + {D - 2\over 2} { \pi \over 2 \beta^2 }  = 0\ .
\label{lamgap}
\eeq
We now restrict to the regime  
\beq
{25 \over 64} \gamma^2 \left({D- 2 \over 2}\right)^2 - 2  \pi  {D- 2 \over 2} > 0  \ ,\label{posdis}
\eeq
for which (\ref{lamgap}) admits two real solutions:
\begin{eqnarray}
(\lambda_1^1)^{1/2} &&= {5 \over 16 \beta}\ \gamma\ {D- 2 \over 2}\nonumber \\
 &&+\ {1\over 2 \beta} \sqrt{
{25 \over 64} \gamma^2 \left({D- 2 \over 2}\right)^2 - 2  \pi  {D- 2 \over 2}} \ ,\label{possol}
\\
(\lambda_1^2)^{1/2} &&= {5 \over 16 \beta}\ \gamma\ {D- 2 \over 2} \nonumber \\
&&-\ {1\over 2 \beta} \sqrt{
{25 \over 64} \gamma^2 \left({D- 2 \over 2}\right)^2 - 2  \pi  {D- 2 \over 2}}\ . \label{negsol}
\end{eqnarray}
In both cases $\lambda_1$ is proportional to $(D-2)/ \beta^2$, which justifies neglecting $t$ in
(\ref{quart}) and implies
that the scale $M^2$ must be chosen  proportional to $1/\beta^2$.
Moreover, since the physical string tension is real we are guaranteed that $M^2
> 0$, as required by the stability of our model. Any complex solutions for ${\cal T}$ would have 
been incompatible with the stability of the truncation.

Let us start by analysing the first solution 
(\ref{possol}). 
By inserting (\ref{possol}) in (\ref{gap3}), we obtain the following
equation for $\rho_1$:
\beq 
{1 \over \rho_1} = 1 - {4 \over 5  + \sqrt{25 - {128 \pi \over \gamma^2 {D-2 \over 2}}}}\ .
\label{rhopos}
\eeq
Owing to the condition (\ref{posdis}),
$1/\rho_1$ is positive
and, since $\lambda_1^2$ is real, the squared free energy is also positive:
\begin{eqnarray}
&&F^2(\beta) \equiv {S^2_{\rm Eff} \over R^2} = {1\over \beta^2}\ \left({5 \over 16 }\
\gamma\ {D- 2 \over 2}\right. \nonumber\\ 
&&-\ \left. {1\over 2 } \sqrt{
{25 \over 64} \gamma^2 \left({D- 2 \over 2}\right)^2 - 2  \pi  {D- 2 \over 2}}
\right)^4 \times\nonumber \\
&&\times \left(1 - {4 \over 5  + \sqrt{25 - {128 \pi \over \gamma^2 {D-2 \over 2}}}} \right) \ .
\label{freew}
\end{eqnarray}
In this case the high-temperature behaviour is the same as in QCD, but the sign is wrong, 
exactly as for the rigid string.
There is, however, a crucial difference: (\ref{freew}) is real, while the
squared free energy for the rigid string is imaginary, signalling an instability in the model.

If we now look at the behaviour of $\rho_1$ at low temperatures, below the deconfining transition
\cite{cris2}, we see that $1/\rho_1$ is positive. The deconfining transition is indeed 
determined by the
vanishing of $1/\rho_1$ at $\beta = \beta_{\rm dec}$. In the case of
(\ref{possol}) this means that   $1/\rho_1$ is positive below the Hagedorn transition, touches
zero at $\beta_{\rm dec}$ and remains positive above it. Exactly the same will happen also for 
$F^2$, which is positive below $\beta_{\rm dec}$ , touches
zero at $\beta_{\rm dec}$ and remains positive above it.
This solution thus describes an unphysical ``mirror'' of the low-temperature behaviour of the
confining string, without a real deconfining Hagedorn transition. For this reason we discard it.

Let us now study the solution (\ref{negsol}).
Again, by inserting (\ref{negsol}) in (\ref{gap3}), we obtain for $\rho_1$ the equation:
\beq 
{1 \over \rho_1} = 1 - {4 \over 5  - \sqrt{25 - {128 \pi \over \gamma^2 {D-2 \over 2}}}}\ .
\label{rhoneg}
\eeq
In this case, when 
\beq
\gamma > 4 \sqrt{{\pi \over 3}} \left({D- 2 \over 2}\right)^{-1/2}\ ,
\label{congam}
\eeq
$1/\rho_1$ becomes negative.
The condition (\ref{congam}) is consistent with (\ref{posdis}) and will 
be taken to  fix the values of the range of parameter $\gamma$ that enters in (\ref{condiz}).
We will restrict to those that satisfy (\ref{congam}). 
Since $\rho_0 = 1$ and $\lambda_1$ is real and proportional to $1/\beta^2$,
we obtain the following form of the squared free energy:
\begin{eqnarray}
F^2(\beta) &&= -{1\over \beta^2}\left( {5 \over 16 }\ \gamma\ {D- 2 \over 2}\right.\nonumber \\
 &&-\ \left.{1\over 2 } \sqrt{
{25 \over 64} \gamma^2 \left({D- 2 \over 2}\right)^2 - 2  \pi  {D- 2 \over 2}}
\right)^4\times\nonumber \\
&&\times \left({4 \over 5  - \sqrt{25 - {128 \pi \over \gamma^2 {D-2 \over 2}}}} -1\right) \ .
\label{freec}
\end{eqnarray}
In the range defined by (\ref{congam}) this is {\it negative}. For this solution, thus,
both $1/\rho_1$ and $F^2$ pass from positive values at low temperatures to negative values at high
temperatures, exactly as one would expect for a string model undergoing the Hagedorn transition at an
intermediate temperature. In fact this is also what happens in the rigid
string case, but there, above the Hagedorn transition, there is a second transition above which, at
high temperature, $\lambda_1$ becomes large and essentially imaginary, giving a positive squared free
energy. This second transition is absent in our model.

A consistency check is made to look if the two solutions  (\ref{possol}) and
(\ref{negsol}), together with (\ref{condiz}) and (\ref{congam}) are compatible with the validity range (\ref{valid}) of
our high-temperature approximation.
Ignoring numerical factors and subleading terms in ${D-2 \over2}$, (\ref{valid}) becomes
\beq
t \beta^2 \ll {D -2 \over 2}\ ,
\label{newcon}
\eeq
which is exactly the condition (\ref{cont}).

Let us now  compare the result (\ref{freec}) with the corresponding one for large-$N$ QCD \cite{polc3}: 
\beq
F^2(\beta)_{\rm QCD} = - {2 g^2(\beta) N \over \pi^2 \beta^2}\ ,\label{qcd}
\eeq
where $g^2(\beta)$ is the QCD coupling constant.
First of all let us simplify our result by choosing large values of $\gamma$:
$$
\gamma \gg \sqrt{128 \pi \over 25} \left({D-2 \over 2}\right)^{-1/2} \ .
$$
In this case (\ref{freec}) reduces to 
\beq
F^2(\beta) = - {1\over \beta^2 } {8 \pi^3 \over 125} {D-2\over \gamma^2}\ .
\label{bingo}
\eeq
This corresponds {\it exactly} to the QCD result (\ref{qcd}) with the identifications
\begin{eqnarray}
g^2 &&\propto {1\over \gamma^2} \ , \nonumber \\
N && \propto D - 2 \ . \nonumber
\end{eqnarray}
The weak $\beta$-dependence of the QCD coupling $g^2(\beta)$ can be accommodated in the parameter
$\gamma$. Note that our result is valid at large values of $\gamma$, i.e. small values of $g^2$, as it
should be for QCD at high temperatures \cite{wilczek}. Note also the interesting identification
between the order of the gauge group and the number of transverse space-time dimensions. Moreover,
since the sign of $\lambda_1$ does not change at high temperatures, the field $x_\mu$ is not
unstable. The contrary happens in the rigid string case \cite{polc2}, where the change of
sign of $\lambda_1$ gives rise to a world-sheet instability.

We thank C. Angelantonj for useful discussions.

\end{document}